\documentclass[preprint,superscriptaddress,prd,showpacs]{revtex4}
\usepackage{amssymb,amsmath,graphicx,amscd,simplewick,xcolor}

\begin{document}

\title{Vacuum Radiation Pressure Fluctuations and Barrier Penetration}

\author{Haiyun Huang}
\email{haiyun.huang@tufts.edu}
\affiliation{Institute of Cosmology, Department of Physics and Astronomy \\
Tufts University, Medford, Massachusetts 02155, USA}

\author{L. H. Ford}
\email{ford@cosmos.phy.tufts.edu}
\affiliation{Institute of Cosmology, Department of Physics and Astronomy \\
Tufts University, Medford, Massachusetts 02155, USA}

\begin{abstract}
We apply recent results on the probability distribution for quantum stress tensor fluctuations to the problem
of barrier penetration by quantum particles. The probability for large stress tensor fluctuations decreases
relatively slowly with increasing magnitude of the fluctuation, especially when the quantum stress tensor
operator has been averaged over a finite time interval. This can lead to large vacuum radiation pressure
fluctuations on charged or polarizable particles, which can in turn push the particle over a potential barrier.
The rate for this effect depends sensitively upon the details of the time averaging of the  stress tensor
operator, which might be determined by factors such as the shape of the potential. We make some estimates
for the rate of barrier penetration by this mechanism and argue that in some cases this rate can exceed the
rate for quantum tunneling through the barrier. The possibility of observation of this effect is discussed. 
\end{abstract}
 

\maketitle
\baselineskip=24pt	

\section{Introduction}
\label{sec:intro}

In a recent paper~\cite{HF15}, we showed how the one loop radiative correction to potential scattering and to
quantum tunneling may be obtained from simple arguments involving the vacuum fluctuations of the time-averaged
quantized electric field. In particular, the one loop enhancement of the quantum tunneling rate obtained by
Flambaum and  Zelevinsky~\cite{FZ99} may be understood as the vacuum electric field giving the particle an
extra boost to get over the barrier. The effects of vacuum electric field fluctuations on light propagation in nonlinear 
materials were discussed in Refs.~\cite{BDFS14,BDFR16}.

In the present paper, we will discuss the effect of vacuum radiation pressure fluctuations in enhancing tunneling rates.
Here we are dealing with fluctuations of the electromagnetic stress tensor, rather than of the fields themselves. 
The role of classical radiation pressure on electrons and atoms in astrophysics has long been studied~\cite{Page20}.
The variance of the radiation pressure fluctuations in a coherent state, which plays a  role in laser interferometer
detectors of gravity waves, was calculated in Refs.~\cite{Caves80,Caves81,SLB95,WF01}. The variance of the 
time averaged radiation  pressure fluctuations in the vacuum state has been treated by several authors in the context of Casimir force
fluctuations~\cite{SLB95,Barton1,Barton2}. Time averaging will play a crucial role in our analysis as well. The fluctuations
of a quantum stress tensor operator at a single spacetime point are not defined in the sense that all of the moments,
beyond the first moment, of such an operator diverge. In general, time averaging of the quantum stress tensor is needed to yield
finite results for the moments. It is also true that the correlation and $n$-point functions of a stress tensor operator are finite
provided that none of the spacetime points involved are at null separations. The Fourier transform of a correlation function
yields a power spectrum, which can be useful for the study of the variance of the fluctuations. This approach was used
in Refs~\cite{JR92,JR93} to study fluctuations of a mirror in the vacuum.

In the present paper, we will consider the effects of large radiation pressure fluctuations in the vacuum state. By ``large",
we mean fluctuations which are much larger than the root-mean-square value found in calculations of the variance.
The probability distributions for quantum stress tensor vacuum fluctuations have been discussed in Refs.~\cite{FFR10,FFR12,FF15}.
These distributions contain the information needed to go beyond calculations of the variance of the fluctuations, 
a fact which was acknowledged by Barton~\cite{Barton1}. The part of the probability distribution which describes large
fluctuations is determined by the higher moments ($n \gg 2$) of the time averaged operator. Thus approaches which focus
upon the variance or the power spectrum of the fluctuations, such as were used in 
Refs.~\cite{Caves80,Caves81,SLB95,WF01,Barton1,Barton2,JR92,JR93} are not particularly useful for the study of large
fluctuations.
A key result is that the distributions for stress tensor fluctuations fall relatively slowly as the magnitude of the fluctuation increases, 
much more slowly than does the Gaussian distribution which describes time averaged electric field fluctuations. This means that large  
radiation  pressure fluctuations  are not so rare as one might have expected. This is especially the case when the relevant
stress tensor has been averaged over a finite time interval~\cite{FF15},  that is, with an averaging function which is
strictly zero outside of a finite interval. Such an averaging functions may be viewed as describing a measurement made over
a finite time. Here we will explore the possible role of large vacuum radiation pressure fluctuations in pushing a particle over a 
barrier more quickly than it would tunnel through the barrier. 

It is well known that at finite temperature, it is possible for particles to acquire enough energy to fly over a barrier without tunneling,
a process known as thermal activation. The effect we will consider bears some similarities to thermal activation, but can occur at
zero temperature. Our effect is also related to the noise-induced activation studied by Antunes, {\it et al}, in Ref.~\cite{Antunes}.
These authors treat a model of a quantum particle in a double well potential which is linearly coupled to a bath of quantum
oscillators. They find a form of activation at zero temperature which can be ascribed to the quantum fluctuations of the oscillator
bath. A key difference between the model of  Ref.~\cite{Antunes} and that in the present paper is that we assume the particle
to be coupled quadratically to the quantized electromagnetic field through the stress tensor. This leads to the possibility of large,
non-Gaussian fluctuations.

The outline of this paper is as follows: The results of Ref.~\cite{FF15} on probability distributions will be summarized in
Sec.~\ref{sec:prob} and extended to the specific case of electromagnetic radiation pressure fluctuations. The effects
of vacuum radiation pressure fluctuations on barrier penetration by charged particles will be examined in Sec.~\ref{sec:hop}. 
Estimates of the  magnitude of this effect will be given, and the conditions under which it can dominate quantum tunneling 
will be discussed. The possible role of radiation pressure fluctuations in nuclear fusion will be treated in Sec.~\ref{sec:nuclear}.
The effect of radiation pressure fluctuations on polarizable, uncharged, particles will be discussed in Sec.~\ref{sec:polar}.
Section~\ref{sec:sum} summarizes and discusses the main results of the paper.

Units in which $\hbar=c=1$, and Lorentz-Heaviside units for electromagnetic quantities will be used unless otherwise noted.

\section{Probability of Large Stress Tensor Fluctuations}
\label{sec:prob}

In this section, we first review previous results on the probability distribution function for
quantum stress tensor fluctuations, and then apply these results to the specific case of
vacuum pressure fluctuations of the quantized electromagnetic field.

\subsection{Finite Duration Measurements and the Probability Distribution}
\label{sec:finite}

Here we summarize the key results of Ref.~\cite{FF15} which will be needed in the present paper.
Let $Q(t)$ be an operator which is a quadratic function of a free field operator, and define its time
average with respect to $f(t)$ by
\begin{equation}
T = \int_{-\infty}^\infty Q(t)\, f(t)\, dt \,,
\end{equation} 
where
\begin{equation}
\int_{-\infty}^\infty  f(t)\, dt = 1 \,.
\end{equation} 
In general, it is the time average, $T$, rather than the local operator, $Q$, which is observable in the sense that one may assign
a well defined probability distribution to $T$, but not to $Q$.
The key idea is that measurements of a quantum stress tensor which occur in a finite time interval should be described by
a sampling function of time, $f(t)$, which is smooth and has compact support. Thus $f(t)$ is taken to be a $C^\infty$,
but non-analytic,
function which is strictly zero outside of a finite time interval whose width is approximately $\tau$. The Fourier
transform of such a function will have an asymptotic form for large argument which falls faster than any power,
but more slowly than an exponential function. Define the Fourier transform by 
\begin{equation}
\hat{f}(\omega) =\int_{-\infty}^\infty dt \,{\rm e}^{-i\omega t} f(t)\,.
\label{eq:Fourier}
\end{equation}
A useful set of compactly supported sampling functions is defined by
\begin{equation}
\hat{f}(\omega) =  {\rm e}^{-|\omega|^\alpha} \,,
\label{eq:alpha}
\end{equation}
where $0 < \alpha < 1$. (Units in which $\tau =1$, following the notation in Ref.~\cite{FF15}, are adopted temporarily.
Later, we return to general units for  $\tau $ when needed for clarity.)
The corresponding functions of time,  $f(t)$, are expressible in terms of Fox H-functions~\cite{Schneider,H-fnts}.
For our purposes, we only require that Eq.~(\ref{eq:alpha}) hold asymptotically for $\omega \gg 1$. 
This will be sufficient to give the switching behavior which we now discuss. We will also require that $\hat{f}(\omega) \geq 0$.
We can arrange for the initial switch-on of $f(t)$, to occur at $t=0$. In this case, the functional form of $f(t)$
as $t \rightarrow 0^+$ is
\begin{equation}
f(t) \sim  t^{-\mu} {\rm e}^{-w\, t^{-\nu}}\,,
\label{eq:switch-on}
\end{equation}
where 
\begin{equation}
\nu = \frac{\alpha}{1-\alpha} \,,
\label{eq:nu}
\end{equation}
\begin{equation}
\mu =  \frac{2-\alpha}{2(1-\alpha)} \,,
\label{eq:mu}
\end{equation}
and
\begin{equation}
w = (1-\alpha)\, \alpha^{\alpha/(1-\alpha)}  \,.
\label{eq:w}
\end{equation}
The switch-off at the end of the finite interval will have the same functional form. The parameter $\alpha$ describes
both the rate of decrease of $\hat{f}(\omega)$, and the behavior of  $f(t)$ at the switch-on and switch-off. A simple
electrical circuit which has a switch-on corresponding to $\alpha = 1/2$ was described in Ref.~\cite{FF15}. In this case,
$f(t) \propto t^{-3/2} \,{\rm e}^{-1/(4t)}$ as $t \rightarrow 0^+$.

The asymptotic form of the Fourier transform, $\hat{f}(\omega)$, determines the rate of growth of the moments of the sampled
stress tensor and in turn, the probability for large fluctuations. Let $T$ be a normal-ordered quadratic operator which 
has been averaged with the sampling function $f(t)$, and define its moments by
\begin{equation}
\mu_n = \langle0| T^n |0\rangle\,.
\end{equation}
We express $T$ in a mode sum of creation and annihilation operators as
\begin{equation}
T = \sum_{i\, j} (A_{i j}\, a^\dagger_i \,a_j + B_{i j}\, a_i \,a_j + 
B^*_{i j} \, a^\dagger_i \,a^\dagger_j ) \,,
\label{eq:T}
\end{equation}
where the coordinate space mode functions are assumed to be plane waves proportional to ${\rm e}^{-i\omega t}$.
Now $\mu_n$ may be expressed as a sum of $n$-th degree polynomials in the coefficients $A_{i j}$ and $B_{i j}$.
These coefficients have the functional forms 
\begin{equation}
A_{i j} \propto (\omega_i \omega_j)^{(p-2)/2}\, \hat{f}(\omega_i -\omega_j )\,,
\end{equation}
and
\begin{equation}
B_{i j} \propto (\omega_i \omega_j)^{(p-2)/2}\, \hat{f}(\omega_i +\omega_j )\,,
\end{equation}
where $p$ is an integer determined by the dimensions of the operator $T$. In the case of stress tensor operators,
which will be our primary concern, $p=3$. However, we will consider the possibility of larger values of $p$ in 
Sec.~\ref{sec:polar}.

It was argued in  Ref.~\cite{FF15} that there is one term in the expression for $\mu_n$ which dominates for $n \gg 1$.
This term is
\begin{equation}
M_n = 4 \sum_{j_1 \cdots j_n} B_{j_1 j_2}\, A_{j_2 j_3}\, A_{j_3 j_4} 
\cdots  A_{j_{n-1} j_n}\,B^*_{j_n j_1}\,.
\label{eq:dom}
\end{equation}
The dominance of this term can be understood as arising from the relative minus sign in the argument of the $\hat{f}$ factor 
in $A_{i j}$, as compared to that in $B_{i j}$. The dominant term contains the maximum number of factors of $A_{i j}$,
which fall more slowly with increasing $\omega_i$. In any case, $M_n < \mu_n$ as all of the terms neglected in
$M_n$ are positive, because $\hat{f}(\omega) \geq 0$. Thus $M_n$ gives a lower bound on the exact moments. 
This will in turn give a lower bound on 
the probability of large fluctuations. In the case where $T$ is a time average of $:\dot{\varphi}^2:$, where
$\varphi$ is the massless scalar field, 
 \begin{equation}
M_n = k_n \int_0^\infty d\omega_1 \cdots d\omega_n (\omega_1 \cdots \omega_n)^p\,
\hat{f}(\omega_1 +\omega_2 ) \hat{f}(\omega_2 - \omega_3 ) \cdots
\hat{f}(\omega_{n-1} - \omega_n ) \hat{f}(\omega_n + \omega_1 ) \,,
\label{eq:Mn} 
\end{equation} 
where
\begin{equation}
k_n = \frac{1}{(2 \pi^2)^n}\,
\label{eq:cn}
\end{equation}
and $p=3$. For $n\gg 1$, the asymptotic form of $M_n$ becomes
\begin{equation}
M_n \simeq k_n\, [2\pi f(0)]^{n-2}\,  \frac{p! [(n-1)p]!}{(np+1)!} \;  
 \int_0^\infty du\, \hat{f}^2(u) \, u^{np+1} \, ,
 \label{eq:Ma1}
\end{equation}
and if $\hat{f}$ has the form given in Eq.~(\ref{eq:alpha}), we have
\begin{equation}
M_n \simeq k_n\, [2\pi f(0)]^{n-2}\,  \frac{p! [(n-1)p]!}{\alpha (np+1)! \, 2^{(np+2)/\alpha}} \; 
 \Gamma\left[\frac{(np+2)}{\alpha}\right]\,.
 \label{eq:Ma}
\end{equation}
The last factor in this expression reveals that for large $n$, the moments grow as $ (p n/\alpha)!$.

This rapid rate of growth of the moments leads to a slow decrease in the tail of the probability distribution. Now 
return to arbitrary units for the sampling time $\tau$ and define the dimensionless variable $x = T\, \tau^{p+1}$. 
Let $P(x)$ be the probability distribution describing the probability of finding various value of $T$ in a measurement.
As explained in Refs.~\cite{FFR10,FFR12}, this probability distribution has a lower bound at the quantum inequality
bound on expectation values of $T$ in an arbitrary state, $x= -x_0 <0$, but no upper bound, so 
\begin{equation}
\int_{-x_0}^\infty P(x)\, dx =1\,.
\end{equation}
The asymptotic form for $P(x)$ for  large $x$ may be written as
\begin{equation}
P(x)  \sim c_0 \,x^b \, {\rm e}^{-a x^c} \,.
\label{eq:tail}
\end{equation}
The constants $c_0$, $a$, $b$, and $c$ may be determined from Eq.~(\ref{eq:Ma}) to be~\cite{FF15} 
 \begin{equation}
c = \frac{\alpha}{p}\,,
\label{eq:c}
\end{equation}
\begin{equation}
b = c\, \left(\frac{2}{\alpha} -p-1\right) -1 
= \frac{2-\alpha}{p} -(\alpha+1)\,,
\label{eq:b}
\end{equation}
\begin{equation}
a = 2\,  [2\pi f(0) B]^{-\alpha/p}\,,
\label{eq:a}
\end{equation}
and
\begin{equation}
c_0 = c\, a^{(b+1)/c}\, B_0\, p! \, \alpha^{-(p+2)} \, 2 ^{-(2/\alpha)}\,
{[2\pi f(0)]^{-2}}\,.
\label{eq:c0}
\end{equation}
Here the constants $B_0$ and $B$ are defined by
\begin{equation}
k_n = B_0 \, B^n \,.
\end{equation}
Thus for the case of $:\dot{\varphi}^2:$, we have $B_0 =1$ and $B=1/(2 \pi^2)$.

Because the moments $\mu_n$ grow faster than $n!$ as $n \rightarrow \infty$, the probability distribution $P(x)$
cannot be uniquely determined by its moments. However, the average behavior of the asymptotic form in
Eq.~(\ref{eq:tail}) can be inferred from the rate of growth of the moments, as was discussed in Refs.~\cite{FFR12,FF15}. 
 It is of interest to seek alternative derivations of the vacuum stress tensor probability distribution, $P(x)$. One
 possibility is numerical diagonalization in a modified theory with a finite number of degrees of freedom. This
 possibility is under investigation. It may also be possible to apply functional approaches, such as the Schwinger-Keldysh, or 
 closed time path method. However, so far this type of approach has been used primarily in perturbative treatments
 and would need to be extended to apply to non-perturbative problems such as that of the probability distribution.

\subsection{Radiation Pressure Fluctuations}
\label{sec:RPF}

Now we wish to apply the results summarized in the previous subsection to the case of vacuum electromagnetic 
radiation pressure fluctuations. These are fluctuations of the time averaged energy or momentum flux
components of the electromagnetic stress tensor. Consider the momentum flux in the $z$-direction 
\begin{equation}
T^{tz} = (\mathbf{E} \times \mathbf{B})^z = E^x\, B^y -E^y\, B^x  \,,
\end{equation}
where $\mathbf{E}$ and $\mathbf{B}$ are the quantized electric and magnetic field operators, respectively.
Let $S^z$ be the momentum flux sampled with $f(t)$
 \begin{equation}
 S^z = \int_{-\infty}^\infty T^{t z}(t,{\bf x})\, f(t)\, dt\,,
 \end{equation}
 where the sampling is in time at a fixed spatial location. Note that $T^{t z}$, and hence $S^z$ are automatically
 normal ordered, as   $\langle0|T^{t z}|0\rangle =0$. The $n$-th moment of $S^z$ is
 \begin{equation}
 \mu_n =  \langle0| (S^z)^n|0\rangle =  
 \int_{-\infty}^\infty dt_1\, f(t_1) \int_{-\infty}^\infty dt_2 \, f(t_2) \cdots  \int_{-\infty}^\infty dt_n \, f(t_n) \;
  \langle0|T^{t z}_1\,   T^{t z}_2\, \cdots T^{t z}_n\,    |0\rangle\, ,
 \end{equation}
 where $T^{t z}_j = T^{t z}(t_j,{\bf x})$. When $n \gg 1$, we expect $\mu_n \sim M_n \sim C_n$, where $C_n$
 is the $n$-th connected moment. 
 
 We expect the high moments of the time averages of both $T^{t z}$ and of $:\dot{\varphi}^2:$, to be of the form of
 Eq.~(\ref{eq:Ma}) with $p=3$, but with different values for the constants $k_n$. We may relate $k_n(T^{t z})$
 to $k_n(:\dot{\varphi}^2:)$, the latter of which are given by Eq.~(\ref{eq:cn}), using a variation of the argument in
 Sec.~IIIB of Ref.~\cite{FFR12}. The connected moments of $:\dot{\varphi}^2:$ may be expressed as a sum of the
 possible connected contractions of the form
\begin{equation}
\contraction{}{\dot{\varphi_1}}{\dot{\varphi_1}}{\dot{\varphi_2}}
\bcontraction{\dot{\varphi_1}}{\dot{\varphi_1}}{\dot{\varphi_2}\dot{\varphi_2}}{\dot{\varphi_3}}
\contraction{\dot{\varphi_1}\dot{\varphi_1}\dot{\varphi_2}}{\dot{\varphi_2}}{\dot{\varphi_3}\dot{\varphi_3}\cdots\dot{\varphi_n}}
{\dot{\varphi_n}}
\bcontraction{\dot{\varphi_1}\dot{\varphi_1}\dot{\varphi_2}\dot{\varphi_2}\dot{\varphi_3}}{\dot{\varphi_3}}{\cdots}{\dot{\varphi_n}}
\dot{\varphi_1}\dot{\varphi_1}\dot{\varphi_2}\dot{\varphi_2}\dot{\varphi_3}\dot{\varphi_3}\cdots\dot{\varphi_n}\dot{\varphi_n} \,,
\end{equation}
where the subscripts label operators at different spacetime points.
 Here the contraction of the form
 \begin{equation}
\contraction{}{\dot{\varphi_i}}{\cdots}{\dot{\varphi_j}}
\dot{\varphi_i}\cdots\dot{\varphi_j}
\end{equation}
contributes a factor of $\langle \dot{\varphi_i} \, \dot{\varphi_j} \rangle$ in the expression for $C_n(\dot{\varphi}^2)$.
 The number of terms in $C_n(\dot{\varphi}^2)$ may be counted as follows: The first operator to contact has $2(n-1)$
 possible partners with which it may be contracted. After this is done, the next operator has $2(n-2)$ possible partners.
 Thus the total number of terms will be
 \begin{equation}
 [2(n-1)][2(n-2)] \cdots 2 = 2^{n-1} \, (n-1)!\,.
 \end{equation}
 The corresponding calculation for the $n$-th connected moment of $S^z$, $C_n(S_z)$, will involve
  \begin{equation}
 \langle  (E^x\, B^y -E^y\, B^x)_1 \, (E^x\, B^y -E^y\, B^x)_2 \cdots (E^x\, B^y -E^y\, B^x)_n \rangle\,.
  \end{equation}
  The contractions of the electric and magnetic field operators are related to those for $\dot{\varphi}$ by the
  relations
  \begin{equation}
 \langle E_i(t) \,E_j(t')\rangle = \langle B_i(t) \,B_j(t')\rangle =
\frac{2}{3}\, \delta_{ij}\, \langle \dot{\varphi}(t) \dot{\varphi}(t')\rangle \,,
\label{eq:2pt}
\end{equation}
and 
 \begin{equation}
 \langle E_i(t) \,B_j(t')\rangle = 0\,,
 \end{equation}
 where all operators are at the same spatial point. This means that $E^x_1$ can only contract with other $E^x$
 operators, etc. Thus  $E^x_1$ has $n-1$ possible contractions, and $B^y_1$ can only contract with other $B^y$
 operators whose associated $E^x$ operator is still uncontracted, as otherwise a disconnected moment would
 result. This  leads to $n-2$ possibilities. The next $E^x$ operator has $n-3$ possibilities, ect. Thus a total of
 $(n-1)!$ terms arise from $E^x\, B^y$, and an equal number from   $E^y\, B^x$, leading to a total of $2(n-1)!$
 terms in $C_n(S_z)$. Equation~(\ref{eq:2pt}) tells us that each contraction of electromagnetic field operators
 contributes a factor of $2/3$ to $C_n(S_z)$, compared to the contribution of a $\dot{\varphi}$ contraction to
 $C_n(\dot{\varphi}^2)$. Thus, we may write
 \begin{equation}
 k_n(S_z) = \left(\frac{2}{3} \right)^n \; \frac{2 (n-1)!}{2^{n-1} \, (n-1)!} \; k_n(\dot{\varphi}^2) = \frac{4}{(6 \pi^2)^n}\,,
 \end{equation}
 where 
 $k_n(\dot{\varphi}^2)$ is given by Eq.~(\ref{eq:cn}). This leads to
 \begin{equation}
 B_0 =4 \quad {\rm and} \quad B= \frac{1}{6 \pi^2}
 \label{eq:BB0}
  \end{equation}
  for $S^z$. This result may also be derived by an alternative argument which involves direct evaluation of the
  vacuum expectation value of a product of $S^z$ operators.
  
  As $p=3$ for $T^{tz}$, and hence for $S^z$, the probability distribution $P(x)$ is a function of $x = \tau^4\, S^z$.
  However, unlike the case of operators such as $\dot{\varphi}^2$ or the energy density, there is no lower bound,
  and the distribution is symmetric $P(-x) = P(x)$. The normalization becomes
  \begin{equation}
\int_{-\infty}^\infty P(x)\, dx =1\,.
\end{equation}
The asymptotic form for $|x| \gg 1$ is still given by Eq.~(\ref{eq:tail}), and the constants $c$, $b$, and $a$ are given
by Eqs.~(\ref{eq:c}), (\ref{eq:b}), and (\ref{eq:a}), respectively, with $p=3$ and $B$ as in Eq.~(\ref{eq:BB0}).
However, the constant $c_0$ is now one-half of that given by  Eq.~(\ref{eq:c0}). The values of the parameters in
the tail of the radiation pressure probability distribution become
 \begin{equation}
c = \frac{\alpha}{3}\,,
\label{eq:c-rp}
\end{equation}
\begin{equation}
b  = -\frac{4 \alpha+1}{3} \,,
\label{eq:b-rp}
\end{equation}
\begin{equation}
a = 2\,  \left[\frac{f(0)}{ 3 \pi} \right]^{-\alpha/3}\,,
\label{eq:a-rp}
\end{equation}
and
\begin{equation}
c_0 = \frac{1}{4 \alpha^4}\, \left[\frac{f(0)}{ 3 \pi} \right]^{2(2\alpha-1)/3}\,
{[2\pi f(0)]^{-2}}\,.
\label{eq:c0-rp}
\end{equation}

 \subsection{Cumulative Probability Distribution}
 \label{sec:cum}
 
 Often we are more interested in a cumulative probability distribution, rather than $P(x)$ itself. Define
  \begin{equation}
  P_>(x) = \int_x^\infty P(y)\, dy \,,
  \end{equation} 
 which is the probability to find any value of $y$ with $y \geq x$ in a given measurement. If $x \gg 1$,
 we may use the asymptotic form for $P(x)$ given in Eq.~(\ref{eq:tail}) to find
 \begin{equation}
  P_>(x) \approx \frac{c_0}{a^{2/c}\, c}\; \Gamma\left(\frac{2}{c}, a x^c \right) \approx
  \frac{c_0}{a\, c}\; x^{1+b-c}\, {\rm e}^{-a x^c} = {\rm e}^{-F(x)} \,,
  \label{eq:cumP}
 \end{equation} 
   where $\Gamma(\frac{2}{c}, a x^c)$ is an incomplete gamma function, and
 \begin{equation}
 F(x) = a\, x^c -(1+b-c)\, \ln x - \ln\left(\frac{c_0}{a c} \right) \,.
  \label{eq:F}
  \end{equation}   
  The constants $a$ and $c_0$ depend upon $f(0)$, the value of the sampling function at $t=0$ in $\tau = 1$
  units. Given that $f(t)$ has unit area and characteristic width $\tau$, we expect $f(0)$ to be of order one. Simple
  choices, such as that illustrated in Fig.~4 of Ref.~\cite{FF15}, give a slightly larger value. For the purposes of
  our estimates, we will set $f(0) = \pi/2$. Then the coefficients which appear in Eqs.~(\ref{eq:tail}) and  
  (\ref{eq:cumP}) for $S^z$, depend only upon the parameter $\alpha$, and are listed in Table~\ref{table:coeff}
  for selected values of $\alpha$.
  
  \begin{table}[htbp]
\caption{ Coefficients for the Radiation Pressure Probability Distribution }
\label{table:coeff}
\begin{center}
\begin{tabular}{|c|c|c|c|c|c|c|} \hline
$\alpha$&$c$&$b$&$a$&$c_0$&$1+b-c$& $ \ln\left(\frac{c_0}{a c} \right)$ \\ \hline
$\frac{1}{2}$ & $\frac{1}{6}$ & $-1$ & $2.70$ & $0.0411$ & $-\frac{1}{6}$ & $-2.39$ \\ \hline
$\frac{1}{3}$ & $\frac{1}{9}$ & $-\frac{7}{9} $ & $2.44$ & $0.310$ & $\frac{1}{9}$ & $0.132$ \\ \hline
$\frac{1}{4}$ & $\frac{1}{12}$ & $-\frac{2}{3} $ & $2.32$ & $1.19$ & $\frac{1}{4}$ & $1.82$ \\ \hline

\end{tabular}
\end{center}
\end{table}

 \subsection{Validity of the Worldline Approximation}
 \label{sec:worldline}
 
 The probability distributions treated in Ref.~\cite{FF15} and reviewed earlier in this section involve only
 time averaging, that is, averaging along the worldline of a point particle in inertial motion. However, in
 realistic physical situations, such as those to be discussed in the next section, some averaging in space
 as well may occur. A systematic treatment of the effects of both space and time averaging will appear in
 Ref~\cite{FF16},  including a discussion of the range of validity of the worldline approximation. 
 This discussion will be briefly summarized here. The effect of spatial averaging can be described by
 a spatial sampling function $g({\bf x})$, with three-dimensional Fourier transform $\hat{g}({\bf k})$. Now
 the expressions for the moments, such as Eq.~(\ref{eq:Mn}), will contain factors of $\hat{g}$ in addition 
 to those of $\hat{f}$, and integrations over $d^3 k_j$. Let $s = \ell/\tau$ denote the ratio the characteristic
 scale of the spatial sampling, $\ell$, to the temporal scale, $\tau$, and assume $s \ll1$. 
 In this case, we expect the worldline approximation to hold for
 the lower moments, and hence the inner part of the probability distribution. 
 
This statement can be made more quantitative as follows: For $\omega \alt 1/s$, we have   
 $\hat{g} \approx 1$. (Recall that $\omega$ is dimensionless in $\tau =1$ units.)
 The dominant contribution in $\omega$ to the $n$-th moment comes near the maximum 
 of the integrand in Eq.~(\ref{eq:Ma1}), which is
 \begin{equation}
\omega_n \approx \left( \frac{n}{2\, c} \right)^{1/\alpha}
\label{eq:omega_n}
\end{equation}
if $\hat{f}$ has the form in Eq.~(\ref{eq:alpha}). Thus the worldline approximation gives an accurate estimate
for the $n$-th moment if
 \begin{equation}
n \alt 2c \, s^{-\alpha} \,.
\end{equation}

For $n \gg 1$, we have
\begin{equation}
\mu_n = \int_{-\infty}^\infty x^n \, f(x) \, dx \approx 2 c_0\, \int_0 ^\infty x^{n+b} \, {\rm e}^{-a x^c}\, dx\,,
\end{equation}
 for the case of the momentum flux $S^z$. The dominant contribution to this integral comes near the maximum of
 its integrand,
 \begin{equation}
x_n \approx \left( \frac{n}{a\, c} \right)^{1/c} \,.
\label{eq:xn}
\end{equation}
 We may now combine these results to infer that the worldline result should give a good approximation to $P(x)$
 for 
 \begin{equation}
x \alt \left(\frac{2}{a}\right)   \, s^{-p}\,.
\label{eq:validity0}
\end{equation}
For the case of stress tensors such as $S^z$, where $p=3$ and $a \approx 2$, as illustrated in 
 Table~\ref{table:coeff}, we find that the worldline approximation gives an accurate estimate for $P(x)$ when
 \begin{equation}
x \alt  s^{-3}\,.
\label{eq:validityS}
\end{equation} 
In addition, we need to have $x \gg 1$, so that the asymptotic probability distribution, Eq.~(\ref{eq:tail}) is valid. We will see below that
there is large region where both conditions may be satisfied.

\subsection{Dependence upon the Switching Parameter $\alpha$}
\label{sec:alpha}

A crucial feature of the asymptotic probability distributions given in Eqs.~\eqref{eq:tail} and \eqref{eq:cumP} is the sensitive
dependence upon the parameter  $\alpha$. A small decrease in the value of  $\alpha$ can cause a significant increase in
the probability of a large stress tensor fluctuation. Recall that this parameter was defined in Eq.~\eqref{eq:alpha}, which gives the 
asymptotic behavior of the Fourier transform, $\hat{f}(\omega)$  of a wide class of compactly supported $C^\infty$ sampling functions. 
The Fourier transform of such a function  must fall faster than any power, but slower than an exponential, and Eq.~\eqref{eq:alpha}
describes the simplest class of functions with this behavior. The rate of decrease of $\hat{f}(\omega)$ for large $\omega$
is linked to the switch-on behavior of the sampling function $f(t)$ through Eqs.~\eqref{eq:switch-on}, \eqref{eq:nu}, \eqref{eq:mu},
and \eqref{eq:w}. Recall that if $\hat{f}(\omega)$ is exactly given by Eq.~\eqref{eq:alpha}, then $f(t)$ is a Fox H-function, but we
are considering a broader class of functions for which Eq.~\eqref{eq:alpha} need only hold asymptotically. Our view is that
the specific form of the sampling function should be determined by the details of the physical system. Note that the variance
of the vacuum radiation pressure fluctuations. which was addressed in Refs.~\cite{SLB95,Barton1,Barton2,JR92,JR93}, is much less
sensitive to the details of the sampling function than is the probability of a large fluctuation, which is the topic addressed here. 
Note that Eq.~(\ref{eq:xn}) implies that the probability distribution for a large value of $x \gg1$ is determined by moments of
order
 \begin{equation}
 n \approx a c \, x^c \gg 1\,.
 \end{equation} 
 This reiterates the point made earlier that studies of the variance or the power spectrum are not adequate to understand large
 fluctuations.

 \section{Barrier Hopping}
 \label{sec:hop}
 
\begin{figure}[htbp]
	\centering
		\includegraphics[scale=0.9]{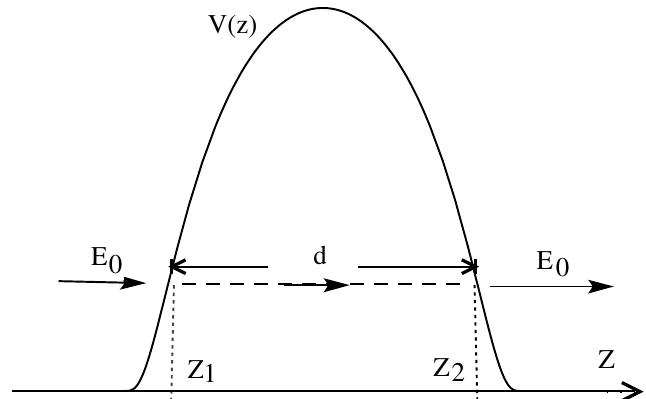}
		\caption{A quantum particle with energy $E_0$ tunnels through a potential barrier $V(z)$. The classical turning points
		are at $z=z_1$ and $z=z_2$. The characteristic spatial width of the barrier is $d = z_2 -z_1$. }
	\label{fig:tunneling}
\end{figure}

 In this section, we will discuss the possible effects of quantum radiation pressure fluctuations on barrier penetration
 by quantum particles. Consider the situation illustrated in Fig.~\ref{fig:tunneling}, where  a particle of mass $m$ and energy $E_0$
 is incident upon a potential barrier $V(z)$, with classical turning points at $z = z_1$ and $z = z_2$, where 
 $E_0 = V(z_1) = V(z_2)$. The probability of quantum tunneling through the barrier is given in the WKB approximation
 by
 \begin{equation}
P_{\rm WKB} = {\rm e}^{-G}\,,
\end{equation}
where
\begin{equation}
G =  2\,\int_{z_1}^{z_2} \sqrt{2\,m \,[V(z) -E_0]} \, dz \,.
\end{equation}
The mean value theorem implies the existence of $z_m$, such that $z_1 \leq z_m \leq z_2$ and
\begin{equation}
G = 2\, \sqrt{2\,m \,[V(z_m) -E_0]} \; d \, ,
\label{eq:Gmv}
\end{equation}
where $d=z_2 - z_1$ is a measure of the spatial width of the barrier. Define a speed $v_1$ by
\begin{equation}
v_1 = \sqrt{2 \,[V(z_m) -E_0]/m} \,, 
\end{equation}
which is the speed of a non-relativistic particle with kinetic energy $V(z_m) -E_0$. Now we can express
$G$ as
\begin{equation}
G = 2\, v_1 \, \left(\frac{d}{\lambda_C}\right) \,,
\label{eq:Gmv2}
\end{equation}
where $\lambda_C = 1/m$ is the reduced Compton wavelength of the particle. Thus, the WKB tunneling probability
decreases as an exponential of the product of speed $v_1$ as a fraction of the speed of light, and of the width
of the barrier as a multiple of the Compton wavelength.   

 \subsection{The Effect of Large Vacuum Radiation Pressure Fluctuations}
 \label{sec:rad-force}
 
\begin{figure}[htbp]
	\centering
		\includegraphics[scale=0.9]{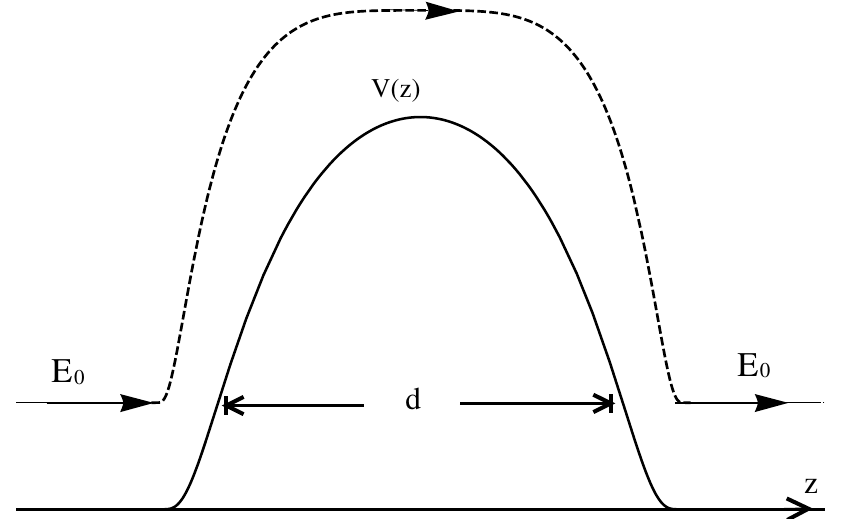}
		\caption{Here the particle temporarily receives extra energy from a quantum radiation pressure fluctuation, which allows it
		to fly over the barrier.  }
	\label{fig:hopping}
\end{figure}

 Now consider the possibility that the particle, while still to the left of the barrier in Fig.~\ref{fig:hopping}, is subjected to a 
 radiation pressure fluctuation in the $+z$ direction. If the magnitude and duration of this fluctuation are
 sufficiently large, it could push the particle over the barrier. Let $\sigma$ be the scattering cross section for
 radiation by the particle, such as the Thompson cross section for a non-relativistic charged particle. The average
 force exerted on the particle by the pressure fluctuation is $\sigma \, S^z$, and the work done if the particle moves
 a distance $d$ to the right during the fluctuation will be
 \begin{equation}
\Delta E = \sigma \, S^z \, d\,.
\end{equation}
 If $\Delta E > V_{\rm max} - E_0$, where $V_{\rm max}$ is the maximum value of the potential, then the 
 particle will fly over the barrier, if the duration of the fluctuation is sufficiently long. Let $v_0$ be the average speed of the 
 particle as it goes over the barrier, and let
  \begin{equation}
\tau = \frac{d}{v_0}
\end{equation}
 be the required duration (in arbitrary units).
 Here we assume that the motion of the particle is non-relativistic, so that the radiation
 pressure in the rest frame of the particle is approximately equal to that in the rest frame of the potential barrier. 
  For the purpose of a rough estimate,  assume that the fluctuation is sufficiently large that 
 $\Delta E$ is at least a few times larger than $V_{z} - E_0$ everywhere and  take 
 $\Delta E \approx \frac{1}{2} \, m\, v_0^2$. Now we may combine the
 above relations to write the dimensionless $x$ as
\begin{equation}
x = \tau^4 \, S^z \approx \frac{m\, d^3}{2\, \sigma \, v_0^2} \,.
\end{equation}
 
 Let the particle have an electric charge of $q$, so $\sigma$ is the Thompson cross section 
 \begin{equation}
\sigma = \sigma_T = \frac{q^4}{6 \, \pi \, m^2} \,.
\label{eq:Thompson}
\end{equation}
Now we can write
\begin{equation}
x = \tau^4 \, S^z \approx \frac{m\, d^3}{2\, \sigma \, v_0^2}\,. 
\label{eq:x}
\end{equation}
 Note that if we hold all other variables fixed and increase $v_0$, and hence $\Delta E$, then $x$ decreases,
 so  $P_>(x)$ typically increases, and the fluctuation becomes more probable. This arises because the factor
 of $1/v_0^4$ coming from $\tau^4$ dominates over the factor of $v_0^2$ in $\Delta E$.
 
  If the cumulative probability, $P_>(x)$ is greater than $P_{\rm WKB}$, or
 \begin{equation}
F(x) < G\,,
\end{equation}
 then the radiation pressure fluctuations will dominate over quantum tunneling. This can occur if $d$ is sufficiently
 large, as $G \propto d$ but $F$ grows more slowly than linearly in $d$. For example, if $\alpha = 1/2$, then 
 $F \propto \sqrt{d}$ for large $d$. For smaller values of $\alpha$, the growth of $F$ with increasing $d$ becomes
 even slower.
 
 Recall that in Sec.~\ref{sec:worldline}, we argued that the validity of the worldline approximation for stress 
 tensor fluctuations requires
 \begin{equation}
x\, s^3 \alt 1\,,
\label{eq:validity}
\end{equation}
 where $s$  is the ratio of the spatial to the temporal averaging scales. In the case of a particle with a scattering
 cross section $\sigma$, we will take the spatial scale to be of order $\sqrt{\sigma}$, and set
 \begin{equation}
s = \frac{\sqrt{\sigma}}{\tau} = \frac{q^2\, \lambda_C}{\sqrt{6\, \pi} \, d}\; v_0 \,.
\end{equation}
 Now Eq.~(\ref{eq:validity}) becomes
 \begin{equation}
x\, s^3 = \frac{q^2}{2 \sqrt{6\, \pi}}\; v_0  \alt 1\,,
\label{eq:validity1}
\end{equation}
where the factors of $\lambda_C$ and of $d$ have canceled.  Let $q = Z\, e$, and recall that 
$e^2 /4\, \pi \approx 1/137$ is the fine structure constant to write Eq.~(\ref{eq:validity1}) as
 \begin{equation}
 \left(\frac{Z}{10}\right)^2 \alt \frac{1}{v_0} \,.
 \label{eq:validity2}
\end{equation}
This condition for the validity of the worldline approximation is generally satisfied for non-relativistic ($v_0 \ll 1$)
elementary particles and smaller nuclei.

 Consider the case of radiation pressure fluctuations on a particle whose charge has a magnitude $e$ such as an 
 electron or proton, so $Z=1$. For the purposes of an estimate, assume that $v_1 \approx v_0$. For  given values
 of $\alpha$ and $v_0$, we may use Eqs.~(\ref{eq:F}), (\ref{eq:Gmv2}), and (\ref{eq:x}), combined with the date
 in Table~\ref{table:coeff}, to find the value of $x$ and hence of $d$ at which $F(x) = G$. A few examples are listed
 in Table~\ref{table:F=G}.  As before, we have estimated the spatial dimension of the worldtube of the particle to be of order
 $\sqrt{\sigma} \approx 0.021\, \lambda_C$, so the ratio of the spatial to the temporal sampling lengths is
 \begin{equation}
 s = \frac{\sqrt{\sigma}}{\tau} \approx  \frac{v_0 \, \lambda_C}{47\, d}\,.
 \end{equation}

  \begin{table}[htbp]
\caption{ Dominance of radiation pressure fluctuations. For given $\alpha$ and $v_0$, this table lists the value
of the width $d$ at which radiation pressure fluctuations begin to dominate over quantum tunneling. }
\label{table:F=G}
\begin{center}
\begin{tabular}{|c|c|c|c|c|c|} \hline
$\alpha$&$v_0$&$G$&$d/\lambda_C $&$x$&$s^{-3}$  \\ \hline
$\frac{1}{2}$ & $0.5$ & $132$ & $132$& $1.0\times 10^{10}$ & $1.9 \times 10^{12}$  \\ \hline
$\frac{1}{2} $  & $0.1$ & $1770$ & $8880$& $ 7.8\times 10^{16}$ & $ 7.3\times 10^{19} $  \\  \hline
$\frac{1}{3}$ & $0.5$  & $12.5 $ & $12.5$ & $8.8 \times 10^6$ & $1.6\times 10^9$   \\ \hline
$ \frac{1}{3}$  & $0.1$ & $54.1$ & $271$ & $2.2 \times 10^{12}$ & $2.1 \times 10^{15}$  \\ \hline
$\frac{1}{4}$ & $0.5$ & $0.64 $ & $0.64$ & $1.2 \times 10^3$ & $2.2 \times 10^5$  \\ \hline
 $ \frac{1}{4}$  & $0.1$    & $3.8 $ & $19$ & $7.6 \times 10^{8}$ & $7.0 \times 10^{11}$  \\ \hline
\end{tabular}
\end{center}
\end{table}

We can draw several inferences from the data in Table~\ref{table:F=G}. First, as the characteristic speed $v_0$
increases, the relative effect of radiation pressure fluctuations increases. This comes from the decrease in the sampling 
time $\tau$ and the corresponding decrease in the parameter $x$. The value $v_0=0.5$ is at the upper limit of 
validity of a non-relativistic treatment, but gives a reasonable order of magnitude estimate of the maximum effect
attainable in this treatment. For  $\alpha= 1/2$,  radiation pressure fluctuations only dominate over quantum tunneling
in a regime where both effects are very small. For example, for $\alpha= 1/2$ and $v_0=0.5$, the probability of 
both effects at the cross over point is of the order of ${\rm e}^{-132}$. However, as $\alpha$ decreases, the 
relative effect of radiation pressure fluctuations increases rapidly. For $\alpha= 1/4$ and $v_0=0.1$, at the point 
that $F=G$, the probability of a particle being kicked over the barrier by a vacuum fluctuation is 
${\rm e}^{-3.8} \approx 0.02$, and for barriers with width $d > 19\, \lambda_C$, radiation pressure fluctuations
will dominate. In all of the cases illustrated, $x\, s^{3} \ll 1$, so the worldline approximation seems to be valid. At the
same time, $x \gg1$, so the asymptotic form, Eq.~(\ref{eq:tail}), of the probability distribution holds.

 \subsection{Sources of the Switching}
 \label{sec:switching}
 
  In this subsection, we will discuss possible physical origins of the switching function, $f(t)$, which averages the
  $T^{tz}$ component of the electromagnetic stress tensor to produce the averaged momentum flux on the particle.
  We are working within the hypothesis that this function must be determined by the details of the physical situation
  or measurement. In the case of a quantum particle impinging upon a potential barrier, one possibility is an
  interplay between the shape of the particle's wavepacket, and the geometry of the barrier. Consider a particle moving
  in one space dimension with wavefunction  $\psi(z,t)$, and hence probability density  $|\psi(z,t)|^2$. It is reasonable to 
  require this to be a compactly supported function of $t$ at fixed $z$, or at least be strictly zero before some specified time.
  This will always be the case if the source of the particle was switched on at a finite time in the past. Although it is
  often convenient to use Gaussian wavepackets, or other functions with infinite tails in both directions, these are 
  idealizations which imply a source in the infinite past. 
  
  Whether the potential $V(z)$ needs to be a compactly supported function of $z$ is less clear. However, it seems reasonable
  to consider such potentials, which describe systems with a finite spatial extent. In this case, we might suppose that the sampling
  of the quantum stress tensor by the particle occurs while the  probability density  $|\psi(z,t)|^2$ and the potential $V(z)$ 
  overlap in space. In this case, $f(t)$ would be zero before the leading edge of the wavepacket reaches the potential, and
  drops again to zero after the wavepacket has split into transmitted and reflected components which have left the region
  where   $V(z) \not=0$. It is also possible to consider potentials of the form $V(t,z)$, with explicit time dependence. 
  Recall that a simple  electrical circuit with switch-on corresponding to $\alpha =1/2$ was discussed in  Ref.~\cite{FF15}.
  
\begin{figure}[htbp]
	\centering
		\includegraphics[scale=0.5]{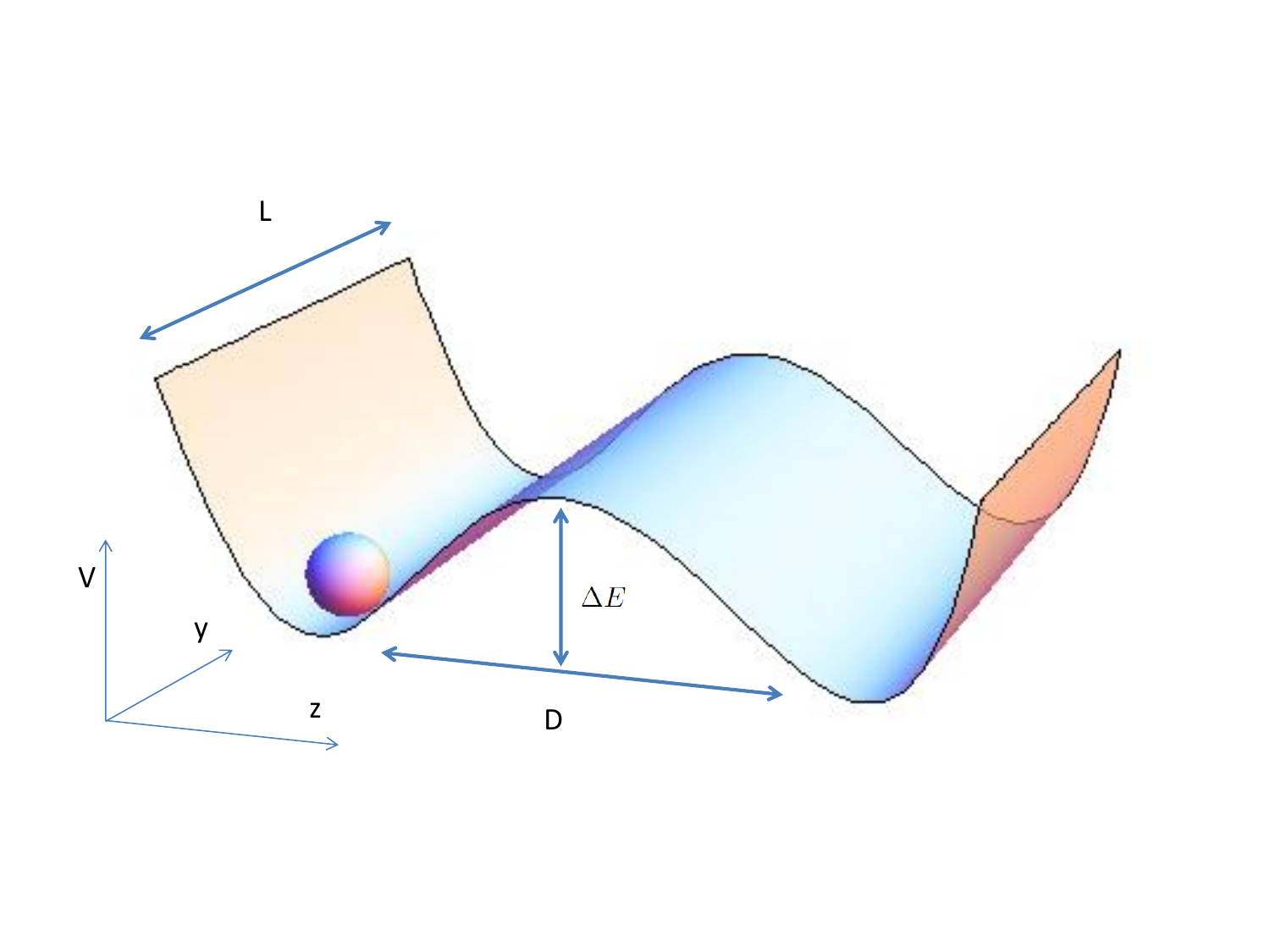}
		\caption{A particle moves along a potential trough in the $y$-direction, which modulates the radation
		pressure fluctuations in the $z$-direction. These fluctuations may in turn push the particle over the barrier.}
	\label{fig:trough}
\end{figure}

  Other possibilities can involve motion in more than one space dimension, as illustrated in Fig.~\ref{fig:trough}.
Here the particle is initially moving in the  $y$-direction in the local minimum of a potential trough on the left. The detailed shape
of the potential as a function of $y$, as well as the shape of the particle wavepacket, define a switching function for
the components of the electromagnetic stress tensor, including $T^{tz}$. This in turn creates an averaged force in the
$+z$-direction, which can cause the particle to jump over the local maximum of the potential to the trough on the right
of the barrier. The temporal switch-on might be modulated by the shape of the potential in the $y$-direction.

\section{Applications to Nuclear Fusion}
\label{sec:nuclear}

An example of barrier penetration by a charged particle arises in nuclear fusion, where a smaller projectile nucleus
must penetrate the Coulomb barrier of a larger target nucleus. For small projectile nuclei, a simple quantum
tunneling  calculation gives reasonable agreement with experiment. However, for larger  projectile nuclei, such as
${}^{16}{\rm O}$, or ${}^{40}{\rm A}$, the simple calculation underestimates the fusion cross section, often by many
orders of magnitude~\cite{Vaz1981,Reisdorf1982}. 
This is usually ascribed to effects such as deformation of the target nucleus. However, we will
explore the possibility that large vacuum radiation pressure fluctuations could be large enough to explain the 
observed cross sections.

\begin{figure}
	\centering
	\includegraphics[width=0.5\textwidth]{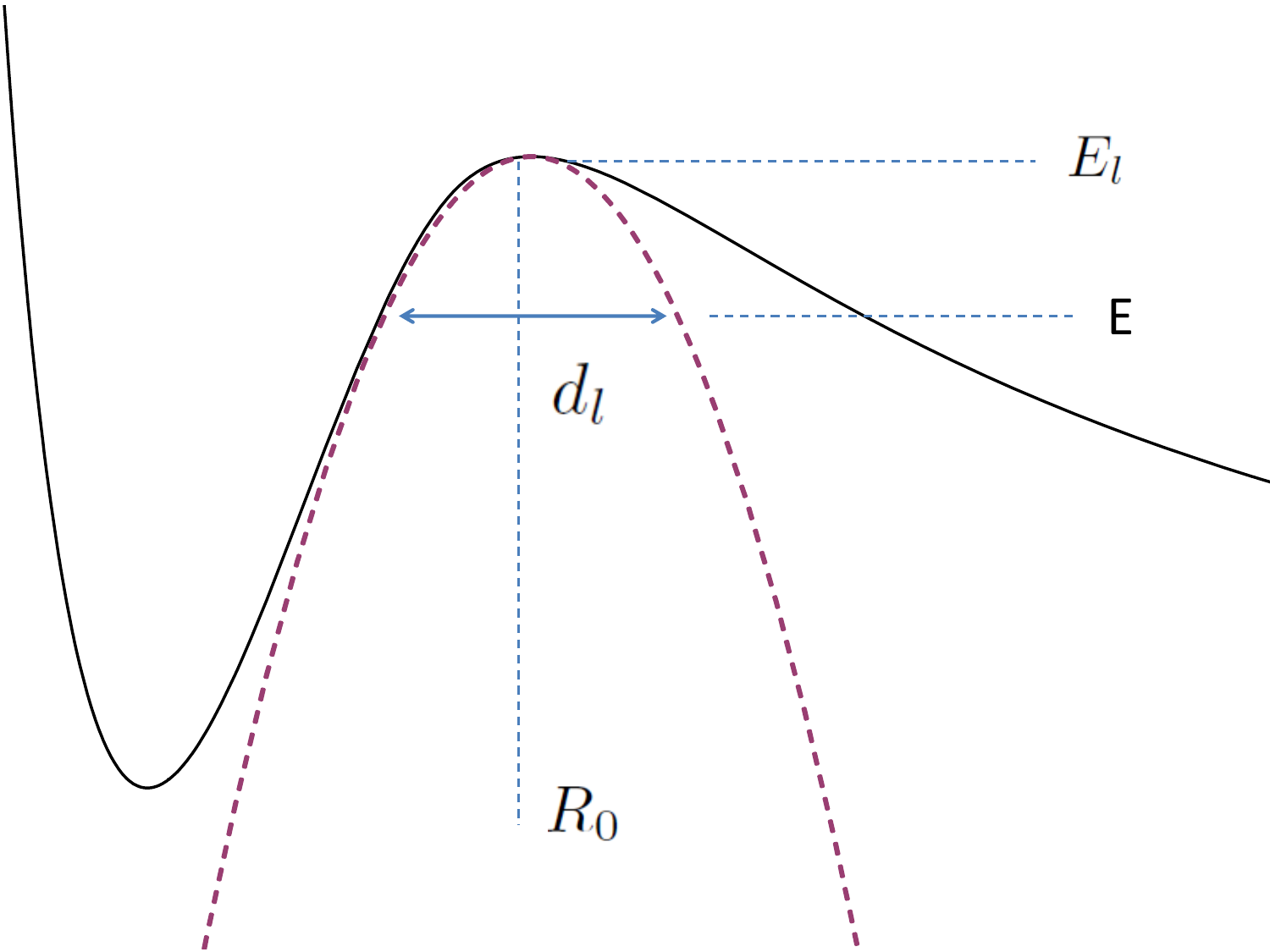}
	\label{fig:nuclear-potential}
	\caption{Sketch of Coulomb barrier for nuclear fusion. The solid curve is the actual potential, which combines
	Couloub repulsion at large separation, and attractive nuclear force at short separation. The dashed curve is the 
	inverted quadratic potential which is tangent to the actual one at the maximum point. Here $d_l$ is the effective width of the
	 barrier at energy $E$.}
\end{figure}

We will consider as an example the fusion of  ${}^{40}{\rm A}$ with ${}^{154}{\rm Sm}$. At a center of mass
energy of $E_{\rm cm} = 113.7\, {\rm MeV}$, the experimentally measured cross section is~\cite{Stokstad1980}
\begin{equation}
\sigma_{\rm exp} = 0.51 \pm 0.10 \, {\rm mb}\,.
\label{eq:exp}
\end{equation}
First, we review the theoretical calculation of the cross section using quantum tunneling in a simple model~\cite{Wong73}. 
Let $\mu$ be the reduced mass
of the system and $k =\sqrt{2 E/\mu}$ be the wavenumber. The cross section may be expressed in a partial
wave expansion as
\begin{equation}
\sigma(E)=\frac{\pi}{k^2}\sum\limits_{l}(2l+1)P_l \,,
\label{eq:sigma}
\end{equation}
where $P_l $ is the transmission probability through the barrier for the $l$-th wave. The potential for this wave
can be modeled by an inverted harmonic oscillator potential
\begin{align}
V_l(r)=-\frac{1}{2}\omega_0^2\mu(r-R_0)^2+E_l\,,
\label{inverted quadratic}
\end{align} 
where
\begin{equation}
E_l=E_0+\frac{l(l+1)}{2 \mu R_0^2}\,.
\end{equation}
Here $\omega_0$, $E_0$, and $R_0$ are parameters which are determined semi-empirically. A fit to
the proximity function given in Ref.~\cite{Santhosh14} leads to the values $E_0=123.4\, {\rm MeV}$, $R_0=12.26\, {\rm fm}$ and
$\omega=4.16\,{\rm MeV}$. This potential models
Coulomb repulsion at larger distances, and nuclear attractive forces at shorter distances, and is illustrated in
Fig.~\ref{fig:nuclear-potential}. The quantum tunneling probability, $P_l $, for this potential is given by the Hill-Wheeler 
formula~\cite{HW53}
\begin{equation}
P_l(E)=\frac{1}{1+\exp[2\pi(E_l-E)/\omega_0]} \,.
\label{eq:Pl}
\end{equation}  
If we evaluate the predicted cross section using Eqs.~(\ref{eq:sigma}) and (\ref{eq:Pl}), with the above choices for the
parameters, the result is
\begin{equation}
\sigma_{HW} \approx 6 \times 10^{-6}  \, {\rm mb} \approx  10^{-5}  \, \sigma_{\rm exp} \,.
\end{equation}
Clearly, the model described above fails badly for below-barrier energies, $E < E_0$. However, it does give
reasonable results for the above-barrier case.

We now explore the hypothesis that the observed cross section in the below-barrier case can be explained by
large vacuum radiation pressure fluctuations, described by the tail of the cumulative probability distribution
given in  Eq.~(\ref{eq:cumP}). Let
\begin{equation}
P_l = P_>(x_l) \approx \frac{c_0}{a c} \,x_l^{1+b-c} \, {\rm e}^{-a x_l^c} \,,
\end{equation}
where
\begin{align}
x_l=\frac{ \mu\, d_l^3}{2\,\sigma_T\, v_0^2}\,.
\end{align}
Here $\sigma_T$ is the Thompson cross section, Eq.~\eqref{eq:Thompson}, and $d_l$ is the width of barrier for the $l$-th 
partial wave, defined by 
\begin{align}
V_l \left(R_0\pm\frac{1}{2}d_l\right) = E \,.
\end{align}
The solutions of this equation are
\begin{align}
d_l= d_0\,[1+\xi\,l(l+1)]^{1/2}\,,
\end{align}
where
\begin{equation}
d_0 = \frac{2}{\omega_0}\; \sqrt{\frac{2(E_0-E)}{\mu}}
\end{equation}
and 
\begin{equation}
\xi = \frac{4}{(\mu\, \omega_0\, R_0\, d_0)^2} \,.
\end{equation}
Define
\begin{equation}
S= \frac{k^2}{\pi} \; \sigma \,,
\end{equation}
so we have
\begin{equation}
S =  \frac{c_0}{a c} \, \sum\limits_{l=0}^{\infty}(2l+1)\, \{x_0[(1+l(l+1)\xi]^{3/2}\}^{1+b-c}\, 
{\rm e}^{-a \{x_0[1+l(l+1)\xi\}^{3/2}]^c} \,.
\label{eq:S}
\end{equation}
For the cases of interest here, this series converges well when about $10^3$ terms are included.

We take the parameters $c$, $b$, $a$, and $c_0$ to be those given by Eqs.~\eqref{eq:c-rp} - \eqref{eq:c0-rp},
with  $f(0) = \pi/2$, and hence functions of $\alpha$ alone. The quantities $x_0$ and $\xi$ are determined by the
parameters specific to the ${}^{40}{\rm A} + {}^{154}{\rm Sm}$ system, and may be expressed as
\begin{equation}
\xi = 4.8 \times 10^{-4}
\end{equation}
and 
\begin{equation}
x_0 = 6.0 \times 10^7 \,.
\end{equation}
In addition, we have $d_0 \approx 2.3 \, {\rm fm}$ in this case. More generally, we can write
\begin{align}
\xi  = 7.4\times10^{-4} \left(\frac{4\, {\rm MeV}}{\omega_0}\right)^2 \left(\frac{32 \, {\rm u}}{\mu}\right)^2
\left(\frac{2\, {\rm fm}}{d_0}\right)^2 \left(\frac{12\, {\rm fm}}{R_0}\right)^2
\end{align}\and
\begin{align}
x_0
&=3.0\times10^{7}\left(\frac{\mu}{{\rm u}}\right)^3 \left(\frac{Z}{18}\right)^2 \left(\frac{d_0}{2\, {\rm fm}}\right)^3
\left(\frac{0.1}{v_0}\right)^2
\end{align}
for any nuclear fusion case, where $Z$ is the atomic number of the incoming nucleus. 

In the case of the ${}^{40}{\rm A} + {}^{154}{\rm Sm}$ system, $Z=18$ and $\mu \approx 32 {\rm u}$. 
At a center of mass energy of $E_{\rm cm} \approx \frac{1}{2} \mu\, v_0^2 \approx 114\, {\rm MeV}$, 
we have $v_0 \approx 0.085$. This leads to $(Z/10)^2 \, v_0 \approx 0.3$. Thus the crtierion for the
validity of the worldline approximation, Eq.~\eqref{eq:validity2}, is satisfied to fair accuracy. This
should be adequate for the order-of-magnitude estimates which we make.

If we replace the sum in Eq.~(\ref{eq:S}) by an integral, $\sum_{l=0}^\infty \rightarrow \int_0^\infty dl$ then
$S \rightarrow S_I$, where $S_I$ may be expressed in terms of an incomplete gamma function:
\begin{equation}
S_I = \frac{2 c_0}{3  c^2 \xi x_0^{2/3}}\, a^{-(5+3b)/(3 c)} \; \Gamma\left(\frac{5+3b-3c}{3c}, a x_0^c \right) \,.
\label{eq:SI}
\end{equation}
If $a x_0^c \gg 1$, we have the asymptotic form
\begin{equation}
S_I  \sim  S_{IA} = \frac{2 c_0}{3 a^2 c^2 \xi }\, x_0^{1+b -2c} \; {\rm e}^{-a x_0^c} \,.
\label{eq:SIA}
\end{equation}

Now we wish to find the value of $\alpha$ which produce a value of $\sigma$ which agrees with the
experimental value, Eq.~(\ref{eq:exp}). This requires $S \approx 2.8$ at $E = 113.7\, {\rm MeV}$. The choices 
which arise from our best estimates of the nuclear parameters, $\xi =4.8 \times 10^{-4} $ and $x_0 = 6.0 \times 10^7$ 
lead to $\alpha \approx 0.27$. The result for $\alpha$ is only weakly sensitive to the values of $\xi$ and $x_0$,
and tend to lie in the range $0.25 \alt \alpha \alt 0.30$, with increases in either $\xi$ or $x_0$ leading to
smaller values for $\alpha$. 
For example, $\xi = 10^{-4}$ and $x_0 = 10^7$ lead to $\alpha \approx 0.30$, while $\xi = 10^{-2}$ and 
$x_0 = 10^8$ lead to $\alpha \approx 0.25$. These results may be obtained from either the sum $S$ or the
integral form $S_I$, which agree very with each other. Thus vacuum radiation pressure fluctuations with $\alpha \alt 0.3$
seem to be large enough to explain the observed cross section.

\section{Radiation Pressure Fluctuations on a Polarizable Particle}
\label{sec:polar}

In this section, we will consider the effects of vacuum radiation pressure fluctuations on an uncharged but electrically
polarizable particle, such as an atom or a neutron. We will assume that the polarizability, $\alpha_0$, is approximately 
independent of frequency.  The Rayleigh scattering cross section for scattering of a monochromatic electromagnetic
wave of angular frequency $\omega$ by such a particle is
\begin{equation}
\sigma_R = \frac{\alpha_0^2}{6 \, \pi} \, \omega^4 \,.
\end{equation}
Thus we can write the force in the $z$-direction on the particle as
\begin{equation}
f^z = \sigma_R\, (\mathbf{E} \times \mathbf{B})^z =  
\frac{\alpha_0^2}{6 \, \pi} \; (\mathbf{\ddot{E}} \times \mathbf{\ddot{B}})^z\,.
\end{equation}
We will assume that the vacuum fluctuations of this force arise from the fluctuations of the operator 
$ (\mathbf{\ddot{E}} \times \mathbf{\ddot{B}})^z$.
More precisely, they arise from the fluctuations of the time averaged operator
\begin{equation}
R^z = \int_{-\infty}^\infty    (\mathbf{\ddot{E}} \times \mathbf{\ddot{B}})^z \;  f(t)\, dt \,,
\end{equation}
where the integrand is evaluated along the world line of the particle. This operator is very similar to the operator $S^z$
treated in Sec.~\ref{sec:RPF}, except for the additional time derivatives, which lead to $p=7$ for $R^z$.

The dimensionless variable, $x$, in the probability distribution $P(x)$ for $R^z$ is now $x = R^z \, \tau^8$. The asymptotic
forms for $P(x)$ and for the cumulative distribution $P_>(x)$ have the forms in Eqs.~(\ref{eq:tail}) and  (\ref{eq:cumP}),
respectively. The numerical constants are determined as before, using $B_0 =4$  and $B= 1/(6 \pi^2)$, as for  $S^z$,
but now using $p=7$. The results are displayed in Table~\ref{table:Rz-coeff}.
\begin{table}[htbp]
\caption{ Coefficients for the Probability Distribution of $R^z$. }
\label{table:Rz-coeff}
\begin{center}
\begin{tabular}{|c|c|c|c|c|c|c|} \hline
$\alpha$&$c$&$b$&$a$&$c_0$&$1+b-c$& $ \ln\left(\frac{c_0}{a c} \right)$ \\ \hline
$\frac{1}{2}$ & $\frac{1}{14}$ & $-\frac{9}{7} $ & $2.27$ & $8.86$ & $-\frac{5}{14}$ & $4.00$ \\ \hline
$\frac{1}{3}$ & $\frac{1}{21}$ & $-\frac{23}{21} $ & $2.18$ & $319.$ & $-\frac{1}{7}$ & $8.03$ \\ \hline
$\frac{1}{4}$ & $\frac{1}{28}$ & $-1 $ & $2.13$ & $3784$ & $-\frac{1}{28}$ & $10.8$ \\ \hline
\end{tabular}
\end{center}
\end{table}
Note that here $c=\alpha/7$, so  $P(x)$ and  $P_>(x)$ decrease very slowly with increasing $x$ and hence
increasing averaged force.

The criterion for the validity of the worldline approximation, Eq.~(\ref{eq:validity0}), now becomes
\begin{equation}
x\, s^7 \alt 1\,,
\end{equation}
where
\begin{equation}
s = \frac{r_0}{\tau}\,,
\end{equation}
and $r_0 = \alpha_0^\frac{1}{3}$ is the characteristic size of the particle. Consider the situation treated in
Sec.~\ref{sec:rad-force}, where the particle can be pushed over a potential barrier by a vacuum force fluctuation.
Here we find
\begin{equation}
x = \frac{3\, \pi \, m d^7}{\alpha_0^2\, v_0^6} \approx \frac{10\, m \, d^7}{r_0^6 \, v_0^6} \,,
\end{equation}
and $s = v_0 \, r_0/d$. Hence $x\, s^7 = 10\, m\, v_0\, r_0$, and the worldline approximation is valid
when
\begin{equation}
v_0 \alt \frac{1}{10\, m\, r_0}\,.
\label{eq:validity-pol}
\end{equation}
This condition is difficult to satisfy for atoms. For the case of a hydrogen atom, for example, we would need
$v_0 \alt 4 \times 10^{-7}$, or $E_0 \alt 8 \times 10^{-8}\, {\rm eV}$, which corresponds to a temperature
below $0.1 \, K$.

The case of the neutron seems more promising, which has a static electric polarizability of 
$\alpha_0 \approx 10^{-3}\,  {\rm fm}^3$ \cite{Schmiedmyer91,Federspiel91,Liebl95}, 
or an spatial size of $r_0 \approx 0.1 \, {\rm fm}$. The validity of
the worldline approximation requires $v_0 \alt 0.2$. Here we will give some estimates for the limiting case
when $v_0 \approx 0.2$ and
\begin{equation}
x \approx 7.8 \times 10^{11} \, \left(\frac{d}{ 1\, {\rm fm}} \right)^7 \,.
\end{equation}
Here
\begin{equation}
G \approx 2\, \left(\frac{d}{ 1\, {\rm fm}} \right)
\end{equation}
and $F$ has the form in Eq.~(\ref{eq:F}), with the coefficients given in Table~\ref{table:Rz-coeff}.
As before, vacuum radiation pressure fluctuations dominate over quantum tunneling when $F < G$.
For the case $\alpha = 1/2$, this begins to occur when $d \approx 80  \,{\rm fm}$, so $F=G \approx 160$,
so the rates due to both effects are very small. When  $\alpha = 1/3$, we have $F=G$ at $d \approx 12.5 \,{\rm fm}$,
corresponding to $P_> = {\rm e}^{-12.5} \approx 3.7 \times 10^{-6}$. 
In the case $\alpha = 1/4$, we find that $F < G$ for all values of $d$,
so the  radiation pressure fluctuation effect dominates. For all values of $\alpha < 1$, for sufficiently large $d$,
we have $F \propto d^\alpha$, and hence growing more slowly than $G$.

\section{Summary and Discussion}
\label{sec:sum}

In this paper, we have explored the hypothesis that large vacuum radiation pressure fluctuations can sometimes
contribute noticeably to barrier penetration by quantum particles with energies below the maximum of the barrier.
This barrier penetration is usually assumed to occur by quantum tunneling, the rate for which decreases exponentially
with increasing barrier height or width. Our analysis is based upon recent results on the vacuum probability distributions
for quantum stress tensor components averaged in time with a class of sampling function with compact support~\cite{FF15}.
We argue that such functions, which vanish outside of a finite time interval, are more realistic descriptions of
physical processes than are functions with tails extending into the infinite past and future. We also suggest that
the choice of the sampling function  should be determined by the details of the physical situation.
Large vacuum radiation pressure fluctuations of the quantized electromagnetic field are
described by a probability distribution which falls more slowly than exponentially, as an exponential of a fractional
power of the sampled pressure. The relatively high probability of large vacuum radiation pressure fluctuations
leads to the possibility that these fluctuations can temporarily give a particle enough energy to fly over the
barrier classically.  The probability of a large fluctuation increases with decreasing time duration
of the sampling function, which measures the time required for the particle to traverse the barrier. Here we have
studied the class of sampling functions reviewed in Sec.~\ref{sec:finite}, which are described by the parameter
$\alpha$, which lies in the range $0< \alpha < 1$. Smaller values of $\alpha$ are associated with a greater
probability of large fluctuations. For non-relativistic charged particles, the force exerted by radiation pressure is proportional
to the Thompson cross section.

Some estimates for the rate of this process were given in Sec.~\ref{sec:rad-force}. It was found that for sufficiently wide
barriers, the vacuum radiation pressure effect can always dominate over usual quantum tunneling. Furthermore, for
sufficiently large incident energies, and hence short sampling times, and for smaller values of $\alpha$, the barrier
penetration rate due to vacuum fluctuation may be large enough to be observable. In Sec.~\ref{sec:nuclear}, we
examined the possible role of  vacuum radiation pressure fluctuations in nuclear fusion, especially heavy ion
projectiles, where the observed fusion cross sections are much larger than predicted by simple barrier tunneling
models. We find that radiation pressure fluctuations with $\alpha \alt 0.3$ could explain the observed cross sections.

In Sec.~\ref{sec:polar}, we turned to force fluctuations on electrically neutral, but polarizable, particles.  Here the classical
force is proportional to the Rayleigh scattering cross section and is proportional to the fourth
 power of the incident wave frequency. We argued that the quantum force fluctuations can be analyzed using the
 probability distribution for the time average of the operator $\mathbf{\ddot{E}} \times \mathbf{\ddot{B}}$, where 
 $\mathbf{E}$ and  $\mathbf{B}$ are the quantized electric and magnetic field operators, respectively. We find the 
 asymptotic form of the probability distribution for this operator averaged with the same class of compactly supported
 sampling functions, and find that it falls even more slowly than does the distribution for averaged stress tensor
 components. We applied  the result to barrier penetration by polarizable particles, using the neutron as an example.
 As in the case of charged particles, it is possible for vacuum force fluctuation effects to dominate over quantum
 tunneling. 

In all cases, the effect is very sensitive to the details of the switching function, particularly to the value of the parameter
$\alpha$. This strong dependence is a new feature of the large vacuum fluctuations being treated in this paper, and 
does not appear when only the variance is considered, as was the case in earlier work~\cite{SLB95,Barton1,Barton2}. 
Our view is that the functional form of the switching function should be determined by the details of the physical
system being studied. Some progress in this direction has been made in the context of nonlinear optical models
for lightcone fluctuations~\cite{BDFS14,BDFR16}, where it was shown that the density profile of a slab of nonlinear
material defines the relevant sampling function for for electric field and squared electric field fluctuations. In the context
of barrier penetration, we have conjectured in Sec.~\ref{sec:switching} that a combination of the shape of the wavepacket 
of the incident particle and the spatial dependence of the barrier potential may also define the relevant sampling function. 
 However, it is not yet clear how to use this information to explicitly determine a value for $\alpha$. 
 This is a topic for future work. In the meantime, we may regard $\alpha$ as an undetermined phenomenological parameter
 which might be possible to determine by experiment.

 \begin{acknowledgments}
 We would like to thank Chris Fewster for valuable discussions.
This work was supported in part by the National Science Foundation under Grants PHY-1506066 and PHY-1607118.
\end{acknowledgments}

\end{document}